\documentclass[12pt]{article}
\usepackage{graphicx}
\usepackage{latexsym}
\usepackage{amsmath,amsfonts,amsthm}

\begin{document}
\title{Low-density series expansions for directed percolation III.  
Some two-dimensional lattices}
\author{Iwan Jensen\thanks{e-mail: I.Jensen@ms.unimelb.edu.au} \\
{\small ARC Centre of Excellence for Mathematics and Statistics of Complex Systems,} \\
{\small Department of Mathematics and Statistics,} \\ 
{\small The University of Melbourne, Victoria 3010, Australia}}
\date{\today}

\maketitle

\begin{abstract}
We use very efficient algorithms to calculate low-density series for  
bond and site percolation on the directed triangular, honeycomb, kagom\'e, 
and $(4.8^2)$ lattices. Analysis of the series yields accurate estimates of 
the critical point $p_c$ and various critical exponents. The exponent estimates 
differ only in the $5^{th}$ digit, thus providing strong numerical evidence 
for the expected universality of the critical exponents for directed percolation 
problems. In addition we also study the non-physical singularities of the series.
\end{abstract}

\section{Introduction \label{sec:intro}}

Percolation is one of the fundamental problems in statistical mechanics
\cite{Essam80a,StaufferPercBook}, and is of great theoretical interest in its own
right as well as being applicable to a wide variety of problems in physics,
biology and many other areas of science. Percolation is commonly formulated 
as a problem on a lattice in which the edges and/or vertices are occupied
(vacant) with probability $p$ ($1-p$). Throughout this paper we shall
refer to {\em occupied} edges and vertices as bonds and sites, respectively,
while edges and vertices refer to the the underlying lattice.  
Nearest neighbour bonds (sites) are said to be connected and clusters are 
sets of connected bonds (sites). Directed percolation (DP) is a 
specialisation to problems in which connections are allowed only along a
preferred direction given by an orientation of the edges of the lattice.

\begin{figure}
\begin{center}
\includegraphics[scale=0.5]{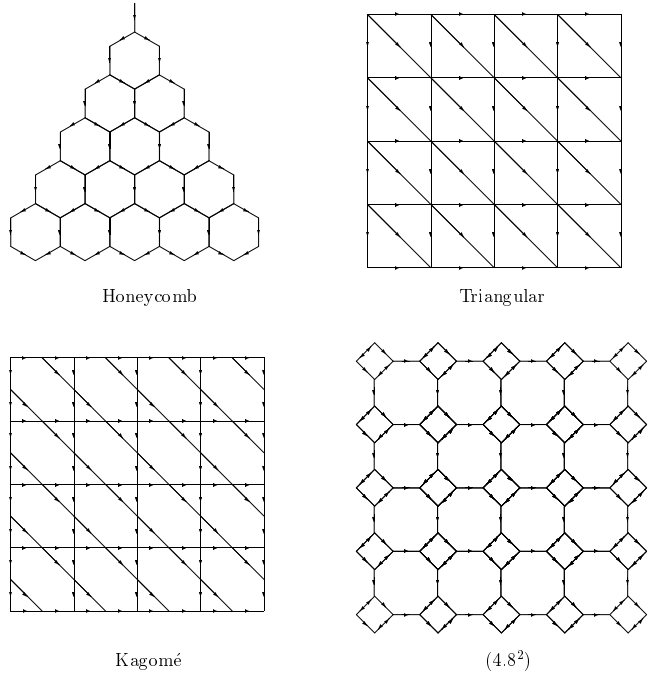}
\end{center}
\caption{\label{fig:lattices}
The directed honeycomb, triangular, kagom\'e, and $(4.8^2)$ lattices.
} 
\end{figure}

We use generalisations of a recently devised and very efficient algorithm 
\cite{IJ99a} to calculate long low-density series for the average cluster 
size and other properties of directed percolation on various two-dimensional 
lattices. In this paper we study bond and site percolation on the directed 
triangular, honeycomb, kagom\'e, and $(4.8^2)$ lattices. 
In figure~\ref{fig:lattices} we show a part of these lattices and the 
orientation of the edges. The $(4.8^2)$ lattice is slightly different from 
the other cases since some edges are bi-directional. In the electrical 
network language some of the edges are occupied by resistors, others are 
occupied by diodes and the remaining edges are insulators (empty). The 
bi-directional edges are required in order to make the problem symmetric. 
Due to the orientation of the remaining edges the lattice has an overall 
preferred direction and the problem is still in the DP universality class.

In the following section we first describe the general graph theoretical
basis of the low-density series expansion technique and then give details 
of how the algorithms have been implemented for each of the lattices studied 
in this paper. In section~\ref{sec:ana} we present the results from our analysis 
of the series, including accurate estimates of the critical point $p_c$ and critical 
exponents and estimates for the location of various
non-physical singularities, as well as the corresponding exponents.

\section{Calculation of low-density series \label{sec:calc}}

In the low-density phase ($p<p_c$) many quantities of interest can be 
derived from the pair-connectedness $C_{t,x}(p)$, which is the probability 
that the vertex at position $x$ is occupied at time $t$ given that the
origin was occupied at $t=0$. The coordinate $t$ measures the distance
from the origin along an axis parallel to the preferred direction, while
the coordinate $x$ measures the distance in the transverse direction. 
Of particular interest are moments of the pair-connectedness

\begin{equation}
  \label{eq:momdef}
  \mu_{m,n}(p) = \sum_t \sum_x t^m x^n C_{t,x}(p)
\end{equation}
\noindent
since they enable us to obtain estimates of the critical point $p_c$ as
well as the critical exponents $\gamma$, $\nu_{\parallel}$, and 
$\nu_{\perp}$. In particular the average cluster size $S(p)=\mu_{0,0}(p)$.
Due to symmetry, moments involving odd powers of $x$ are identically zero.
The remaining moments diverge as $p$ approaches $p_c$ from below

\begin{equation}
  \label{eq:momdiv}
  \mu_{m,n}(p) \propto (p_c-p)^{-(\gamma+m\nu_{\parallel}+n\nu_{\perp})},
  \;\;\;\; p \rightarrow p_c^-
\end{equation}
\noindent

It has been show \cite{AE77} that the pair-connectedness can be expressed 
as a sum over all graphs $g$, which are coverable by a union of directed
paths connecting the origin to the vertex at $(t,x)$,

\begin{equation}
  \label{eq:pairconn}
  C_{t,x}(p)= \sum_g d(g)p^{|g|},
\end{equation}
\noindent
where $|g|$ is the number of bonds in $g$. The weight $d(g)=(-1)^{c(g)}$,
where $c(g)$ is the cyclomatic number of the graph $g$. The restriction
to coverable graphs is very strong and leads to a huge reduction in the
number of graphs that need be counted in order to calculate the 
pair-connectedness. One immediate consequence is that graphs with dangling 
parts make no contribution to $C_{t,x}$ and any contributing graph 
terminates exactly at $(t,x)$. Another way of stating the restriction
is that any vertex with an incoming bond {\em must} have an outgoing
bond unless it is the terminal vertex $(t,x)$.

Any directed path to a site whose parallel distance from the origin is
$t$ contains at least $k(t)$ bonds, where $k(t)$ is lattice dependent. From 
this it immediately follows that if $C_{t,x}$ has been calculated for
$t \leq t_{\rm max}$ then one can determine the moments to order
$k(t_{\rm max})$. One can however do much better as shown in the work
by Essam et al. \cite{EGDB88a}. They used a so-called non-nodal graph
expansion, based on work by Bhatti and Essam \cite{BE84}, to extend the
series to order $2k(t_{\rm max})+1$ on the square and triangular
lattices. A graph $g$ is nodal if it has a vertex (other than the terminal
vertex) through which all paths pass. It is clear that each such nodal
point effectively works as a new origin for the cluster growth, and
we can obviously obtain any coverable graph by concatenating non-nodal
graphs. More precisely we concatenate two coverable graphs $g'$ and
$g''$ by placing the original vertex of one graph on top of the terminal
vertex of the other graph. If $g'$ terminates at $(x',t')$ and $g''$
terminates at $(x'',t'')$ then

\begin{eqnarray}
  \label{eq:concat}
  g & = & g'g'' \;\; \mbox{terminates at} (x'+x'',t'+t''), \nonumber \\
  |g| & = & |g'| + |g''|, \\
  c(g) & =& c(g') + c(g''). \nonumber
\end{eqnarray}
\noindent
Note that the graph consisting of a single bond is non-nodal so all
linear graphs can be obtained by repeated concatenations. This is
the essential idea behind the non-nodal graph expansion, which proceeds
in two principal steps. First we calculate the contribution $C^*_{t,x}$
of non-nodal graphs to the pair-connectedness. Next we use repeated
concatenation operations of $C^*_{t,x}$ to calculate the pair-connectedness
$C_{t,x}$ and from this we finally calculate various moments $\mu_{m,n}(p)$.
One is mainly interested in moments involving only $t$ or $x$ (that is
cases with $n=0$ and/or $m=0$), and in practice we calculate only the
first two non-zero moments. The exclusion of `cross' moments has the
advantage that we need only calculate two-parameter generating functions
$C_t(p)$ and $C_x(p)$. The savings in time obtained by the elimination of
one variable more than compensates for the minor complication of having
to calculate these two functions separately.

The calculation of $C^*_{t,x}$ is done efficiently using transfer-matrix
techniques. This involves drawing a boundary line across a finite
slice of the lattice and then moving the boundary line such that
one adds row after row with each row built up one lattice cell at a time.
The sum over all contributing graphs is calculated as the lattice is
constructed. At any given stage the boundary line cuts through a
number of, say $j$, vertices. There are two possible states (0 or 1) per 
vertex leading to a total of $2^j$ possible boundary configurations. In 
site percolation the different states correspond to the vertex being
empty (0) or occupied (1), while in bond percolation we distinguish
between vertices with (1) and without (0) incoming occupied edges.
The weight of each configuration is given by a polynomial in $p$
truncated at order $N$.

The maximal order $N$ to which the series can be calculated is primarily
limited by available computer memory. While the maximal number of
boundary configurations is $2^j$ it must be emphasised that not all
of these are required because they represent graphs contributing at an
order exceeding $N$. Savings in memory use can be achieved as follows.
Firstly, for each boundary configuration we keep track of the current
minimum number of bonds $N_{\rm min}$ that have already been inserted.
Secondly, we calculate the minimum number of addition bonds $N_{\rm add}$
required to produce a valid non-nodal graph. If $N_{\rm min}+N_{\rm add}>N$
we can discard that configuration because it won't make a contribution
to the pair-connectedness up to the maximal order we are trying to obtain.
$N$, as well as $N_{\rm min}$ and $N_{\rm add}$, depend on the given lattice
and on the specific implementation of the transfer-matrix algorithm and one
naturally tries to find an implementation which is simple yet tends to
maximise $N$. In the following sub-sections we give some details of the
transfer-matrix algorithms and concatenation operations for the case
of directed {\em bond} percolation on the four lattices studied in this
paper. The extension to site percolation is outlined briefly.

\subsection{Honeycomb lattice \label{sec:hccalc}}

Percolation on the directed honeycomb lattice is a simple generalisation
of the directed square lattice problem. The various moments for site percolation 
can be obtained from the square series by the simple substitution 
$p \rightarrow p^2$ \cite{DPB82,EDB82}, and we have thus not investigated
this problem further. Bond percolation on the other hand is closely related 
to the problem of site-bond percolation on the square lattice in which both
sites and bonds are present with probability $p$.

\subsubsection{The transfer-matrix algorithm \label{sec:hctransfer}}

\begin{figure}
\begin{center}
\includegraphics[scale=0.8]{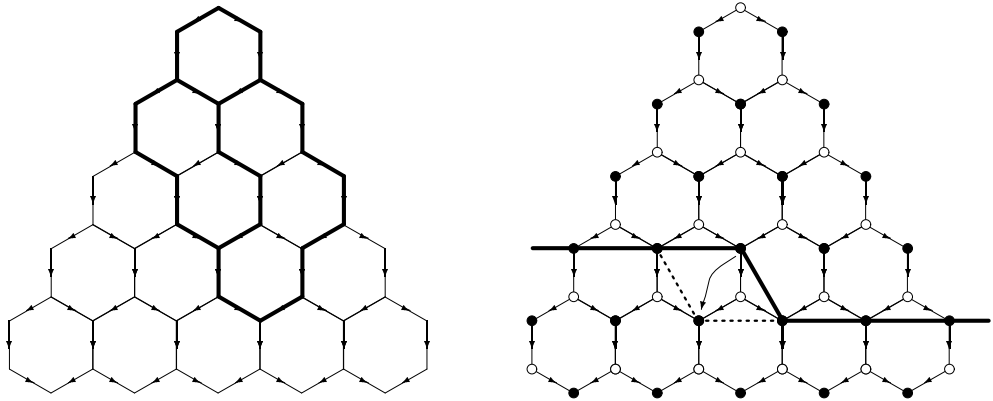}
\end{center}
\caption{\label{fig:hctransfer}
A piece of the directed honeycomb lattice. In the left panel we show an 
example of a non-nodal graph with 24 bonds and 3 cycles. 
In the right panel we show a snapshot of the boundary
line (thick solid) during the transfer-matrix calculation and indicate 
how it is moved to a new position (dotted line) while adding an extra
`cell' with two vertices and three edges to the lattice.
} 
\end{figure}

The algorithm used to calculate the non-nodal contributions $C^*$ to the
pair-connectedness is a simple generalisation of the square lattice 
algorithm \cite{IJ99a}. In figure~\ref{fig:hctransfer} we show a
snapshot of the boundary line (solid) during the transfer matrix 
calculation and indicate how it is moved to a new position (dotted line)
while adding an extra `cell' to the lattice. The updating of the weights, $W$,
associated with the boundary configurations depend only on the states
of the topmost vertex and the neighbours to the left and right in the
row below. The only difference from the square lattice rules is that
when the top vertex is in state `1' we have to insert the extra vertical
bond leading to the middle vertex (indicated by open circles) since only
graphs with no dangling ends make a contribution. Thus we obtain the
honeycomb lattice updating rules from those on the square lattice simply
by inserting an extra factor $p$ on all weights of configuration with the 
topmost vertex in state `1':

\begin{eqnarray}
W(\overline{S}_{1,1}) & = & p^2 W(S_{1,0})+p^2W(S_{1,1})-p^3W(S_{1,1}), \nonumber \\
W(\overline{S}_{0,1}) & = & W(S_{0,1}) + p^2W(S_{1,0}) - p^2W(S_{1,1}),  \\
W(\overline{S}_{1,0}) & = & p^2 W(S_{1,0}), \nonumber \\
W(\overline{S}_{0,0}) & = & W(S_{0,0}), \nonumber
\end{eqnarray}
\noindent
where $S_{i,j}$ is a boundary configuration {\em before} the move with the 
topmost vertex in state $i$ and the rightmost vertex in state $j$, while
similarly $\overline{S}_{i,j}$ is a boundary configuration {\em after} 
the move with the leftmost vertex in state $i$ and the rightmost vertex 
in state $j$. Other aspects of the algorithm are as in \cite{IJ99a},
except that the minimum order to which a configuration can make a
contribution to $C^*_{t,x}$ from row $t'$ is

\begin{equation}
N_{\rm cont} = 2N_{\rm min}+N_1+2t-4t'.
\end{equation}
\noindent
$N_1$ is the number of vertices in state `1' and this factor reflects 
the fact that for 
each vertex in state `1' we have to insert an extra vertical bond.

The calculation was carried out for values of $t$ up to 75 and the series 
were thus obtained to order 301.

\subsubsection{The concatenation operations \label{sec:hcconcat}}

Any graph contributing to $C_{t,x}$ is either a linear graph or can be
broken into non-nodal components connected by linear pieces and possibly
ending with a linear piece. Using standard techniques of generating 
functions we have

\begin{equation}\label{eq:hcconcat}
C_{t,x}=E+LC^*E+LC^*LC^*E+\ldots=\frac{E}{1-LC^*}.
\end{equation}
\noindent
The generating function, $L$, for the linear pieces is easily obtained 
by noting that a linear graph connecting two non-nodal pieces starts 
at a vertex in the left panel of figure~\ref{fig:hctransfer} indicated 
by a closed circle and terminates at a vertex indicated by an open circle.
Thus a linear piece can consist of a single step down (this adds a bond
and increase $t$ by one so it is represented by a factor $pt$) or a
down step followed by repeated instances of either a step to the left or
a step to the right and a step down

\begin{equation}
L=pt+p^3t^3(x+x^{-1})+p^5t^5(x+x^{-1})^2+\ldots = 
   \frac{pt}{1-p^2t^2(x+x^{-1})},
\end{equation}
\noindent
where we put in a weight $x$ ($x^{-1}$) for each step to the right (left).
Similar arguments demonstrate that the generating function, $E$, for the
end piece is 

\begin{equation}
E= \frac{1+pt}{1-p^2t^2(x+x^{-1})}.
\end{equation}

Note that the variables $x$ and $t$ appearing in these generating functions
aren't the same as those appearing in the definition of the moments in 
Eq.~(\ref{eq:momdef}). The relationship is simply that {\em the power} of
$x$ and $t$ in the generating functions are the $x$ and $t$ values which should
be used in calculating the moments.

\subsection{Triangular lattice \label{sec:tricalc}}

On the triangular lattice we calculate the weight of non-nodal graphs
starting in the top left corner and terminating in the bottom right corner
of rectangles of size $w \times l$, as illustrated in the left panel of
figure~\ref{fig:tritransfer}. Such graphs contribute to $C^*_{t,x}$, where
$t=w+l$ and $x=l-w$, and contain at least $t+\max (x,1)$ bonds. Due to
symmetry we need only consider rectangles with $l \geq w$. So if we want
to calculate the series to order $N$ we must sum the contributions from 
all rectangles with widths up to $N/2$ and lengths from $w$ to $N/2$.  

\subsubsection{The transfer-matrix algorithm \label{sec:tritransfer}}

\begin{figure}
\begin{center}
\includegraphics[scale=0.8]{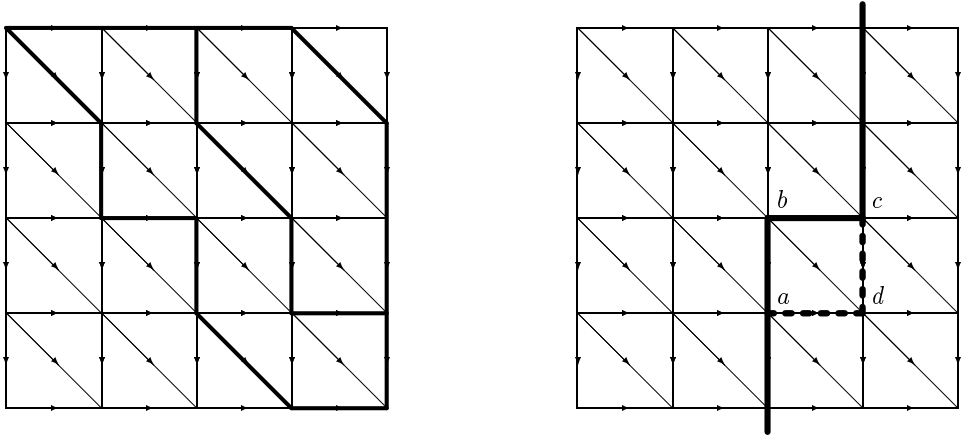}
\end{center}
\caption{\label{fig:tritransfer}
A piece of the directed triangular lattice. The left panel contains an 
example of a non-nodal graph while the right panel illustrates how the 
boundary line is moved  during the transfer-matrix calculation in order
to build up a rectangular slice of the lattice.
} 
\end{figure}

The right panel of figure~\ref{fig:tritransfer} shows how the boundary 
line is moved in order to add an extra cell to the triangular lattice.
As the boundary line is moved we insert bonds emanating from the vertex
in the top left corner of the kink (in figure~\ref{fig:tritransfer} this
vertex is marked $b$), so only if this vertex is in state `1' are any
bonds added. Note that bonds have been inserted only to the left of the
boundary line while those {\em on} and to the right may be inserted in 
subsequent moves. Depending on which bonds are added the states of the
vertices marked $a$ and $c$ can change while the `new' vertex $d$ is 
assigned a state. We use the notation $\overline{S}_{i,j,k}$ ($S_{i,j,k}$)
to indicate a boundary configuration after (before) the move with vertices
$a$, $d$, and $c$ ($a$, $b$, and $c$) in states $i$, $j$, and $k$, 
respectively. The updating rules show how to obtain the weight
$W(\overline{S}_{i,j,k})$ of a given `target' configuration from the
weights of the `source' configurations. Considerations similar to those
for the square lattice case \cite{IJ99a} yield

\begin{eqnarray}\label{eq:tribondupdate}
W(\overline{S}_{1,1,1}) & = & 
        (p-2p^2+p^3)W(S_{1,1,1})+(p^2-p^3)W(S_{0,1,1}) \nonumber\\
  & & + (p^2-p^3)W(S_{1,1,0}) + p^3W(S_{0,1,0}), \nonumber \\
W(\overline{S}_{0,1,1}) & = &   
        (p-p^2)W(S_{0,1,1})+ p^2W(S_{0,1,0}), \nonumber \\
W(\overline{S}_{1,0,1}) & = & 
        (-2p+p^2)W(S_{1,1,1})+(p-p^2)W(S_{0,1,1}) \nonumber\\
  & & + W(S_{1,0,1})+ (p-p^2)W(S_{1,1,0}) + p^2W(S_{0,1,0}), \nonumber \\
W(\overline{S}_{0,0,1}) & = & 
       -pW(S_{0,1,1}) + W(S_{0,0,1})+pW(S_{0,1,0}), \\
W(\overline{S}_{1,1,0}) & = &   
        (p-p^2)W(S_{1,1,0})+ p^2W(S_{0,1,0}),  \nonumber \\
W(\overline{S}_{0,1,0}) & = & pW(S_{0,1,0}), \nonumber \\
W(\overline{S}_{1,0,0}) & = & 
       -pW(S_{1,1,0}) +pW(S_{0,1,0})+ W(S_{1,0,0}), \nonumber \\
W(\overline{S}_{0,0,0}) & = & W(S_{0,0,0}). \nonumber 
\end{eqnarray}
\noindent
 
We shall only briefly indicate how these equations are derived by looking 
at the updating rule for $\overline{S}_{1,1,1}$. The contribution from
$S_{0,1,0}$ is the simplest. Since only the vertex $b$ has an incoming
bond all three bonds have to be inserted in order to yield the target
$\overline{S}_{1,1,1}$ and we thus get the contribution $p^3W(S_{0,1,0})$.
The contribution from $S_{1,1,0}$ arise as follows. Since vertex $c$ is
in state `0' we have to insert the two bonds from $b$ to $c$ and $d$.
The bond from $b$ to $a$ can be either absent or present. If the bond
is present, we note that since vertex $a$ already has an incoming bond
the additional bond results in the formation of an extra cycle and
we have to weight this case with a factor $-p^3$. Thus the total
contribution is $(p^2-p^3)W(S_{1,1,0})$. Due to symmetry an identical
weight is assigned to the contribution from $S_{0,1,1}$. Finally,
the contribution from $S_{1,1,1}$ is derived by noting that the diagonal 
bond from $b$ to $d$ must be inserted while the remaining two bonds can
be either absent or present. If only one of these bonds is inserted we 
add in total two extra bonds and form one extra cycle. When both these 
bonds are inserted we form two extra cycles. All in all we get the 
contribution $(p-2p^2+p^3)W(S_{1,1,1})$. The remaining updating rules
can be derived via similar lines of reasoning.

In this case $N_{\rm add}$ has to be calculated using a simple algorithm.
There are essentially only two contributions. Firstly there is the number
of extra bonds required to connect the top and bottommost sites with
incoming bonds to the first permissible vertex on the bottom row.
Since we have to span at least a $w\times w$ rectangle the first column
at which the two paths can join is $l = \max (w,l'+\delta)$, where $l'$
is the current column position of the boundary and $\delta$ is a small
number depending on the position of top and bottommost sites with
incoming bonds with respect to a possible kink in the boundary line. $\delta$ 
is 2 if there is a kink in the boundary {\em and} the bottommost site is 
{\em above} the kink, 1 if there is a kink in the boundary 
{\em and} the topmost site is {\em above} the kink, 0 otherwise
The connection of the two sites has to be done in such a way as to
ensure there are no nodal vertices, so we are just dealing with two
non-intersecting directed walks joining at $(l,w)$. We shall refer to
these two walks as the outer perimeter. Secondly, some additional bonds
are added in order to connect any intermediate sites to the outer
perimeter. We have to choose the shape of the outer perimeter so as
to minimise the number of extra bonds. In practice this is quite a
simple minimisation problem and it is easy to write the required
algorithm.

We were able to calculate $C^*$ up to width 44 and consequently obtain 
the series to order 90. This is a very substantial improvement on the
previous best of 57 terms \cite{IJ96a}.

\subsubsection{The concatenation operations \label{sec:triconcat}}

The concatenation operations are very similar to the honeycomb case,
except that the end piece and the linear connecting pieces are identical.
So in this case we have

\begin{equation}\label{eq:triconcat}
C_{t,x}=L+LC^*L+LC*LC^*L+\ldots=\frac{L}{1-LC^*}.
\end{equation}
\noindent
The generating function for a  linear piece is 

\begin{equation}
L=   \frac{1}{1-ptx-ptx^{-1}-pt^2},
\end{equation}
\noindent
which represents repeated single steps either to the right, down or
along the diagonal.

\subsubsection{The site problem \label{sec:trisite}}

The transfer matrix algorithm for the site problem is very similar to that 
of the bond case, but of course the weights used in the updating rules
are quite different. The updating rules for the site case are however
easy to derive from the bond rules. All we have to do is change the weights 
in eq.~(\ref{eq:tribondupdate}) so that we count the number
of additional occupied vertices rather than edges. This results in many
cancellations and leads to the following site updating rules:

\begin{eqnarray}\label{eq:trisiteupdate}
W(\overline{S}_{1,1,1}) & = & p^3W(S_{0,1,0}), \nonumber \\
W(\overline{S}_{0,1,1}) & = & p^2W(S_{0,1,0}), \nonumber \\
W(\overline{S}_{1,0,1}) & = & 
        -pW(S_{1,1,1}) + W(S_{1,0,1}) + p^2W(S_{0,1,0}), \nonumber \\
W(\overline{S}_{0,0,1}) & = & 
       -pW(S_{0,1,1}) + W(S_{0,0,1})+pW(S_{0,1,0}), \\
W(\overline{S}_{1,1,0}) & = & p^2W(S_{0,1,0}),  \nonumber \\
W(\overline{S}_{0,1,0}) & = & pW(S_{0,1,0}), \nonumber \\
W(\overline{S}_{1,0,0}) & = & 
       -pW(S_{1,1,0}) +pW(S_{0,1,0})+ W(S_{1,0,0}), \nonumber \\
W(\overline{S}_{0,0,0}) & = & W(S_{0,0,0}) \nonumber 
\end{eqnarray}
\noindent

We calculated $C^*$ up to width 40 and obtained the series to order 82 
as compared to the previous best of 57 terms \cite{IJ96a}.

\subsection{Kagom\'e lattice \label{sec:kagcalc}}

In order to facilitate our calculation of $C_{t,x}$ we divide the vertices
of the kagom\'e lattice into three subsets as indicated in the left panel  
of figure~\ref{fig:kagtransfer}. In the following we shall refer to vertices
indicated by black, shaded, or open circles as being of type $a$, $b$, or
$c$, respectively. The transfer matrix algorithm is designed to calculate
$C^*_{a,a}$, that is the contribution to the pair-connectedness from 
non-nodal graphs starting and terminating on vertices of type $a$. From
this we use concatenation operations to calculate the full pair-connectedness
$C_{t,x}$, starting at a vertex of type $a$ and terminating on a vertex
of any type.

\subsubsection{The transfer-matrix algorithm \label{sec:kagtransfer}}

\begin{figure}
\begin{center}
\includegraphics[scale=0.8]{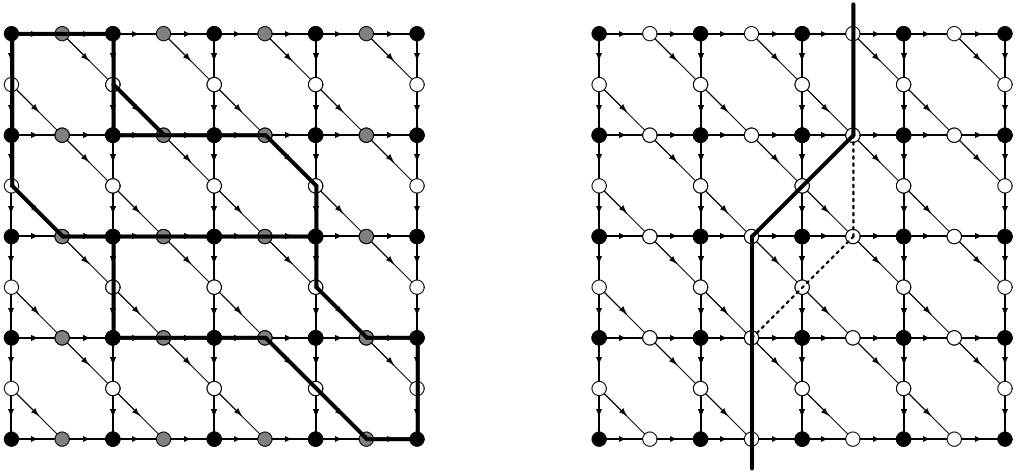}
\end{center}
\caption{\label{fig:kagtransfer}
A piece of the directed kagom\'e lattice. Indicated in the left panel are 
the three different vertex types used in the calculation of the 
pair-connectedness and an example of a non-nodal graph contributing to 
$C^*_{a,a}$. The right panel illustrates how the boundary line is moved  
during the transfer-matrix calculation.
} 
\end{figure}

In this case we calculate the contribution of non-nodal graphs starting in
the top left corner and terminating in the bottom right corner of 
$w \times l$ rectangles, as illustrated in the left panel of
figure~\ref{fig:kagtransfer}. Due to symmetry we need only consider
rectangles with $l \geq w$. These graphs contribute to $C^*_{t,x}$,
where $t=2(l+w)$ and $x=2(l-w)$, and they contain at least 
$t+x+2=4l+2$ bonds. So in a calculation to order $N$ we have to calculate
the contribution from rectangles with width up to $N/4$ and lengths from
$w$ to $N/4$.

\begin{figure}
\begin{center}
\includegraphics[scale=0.8]{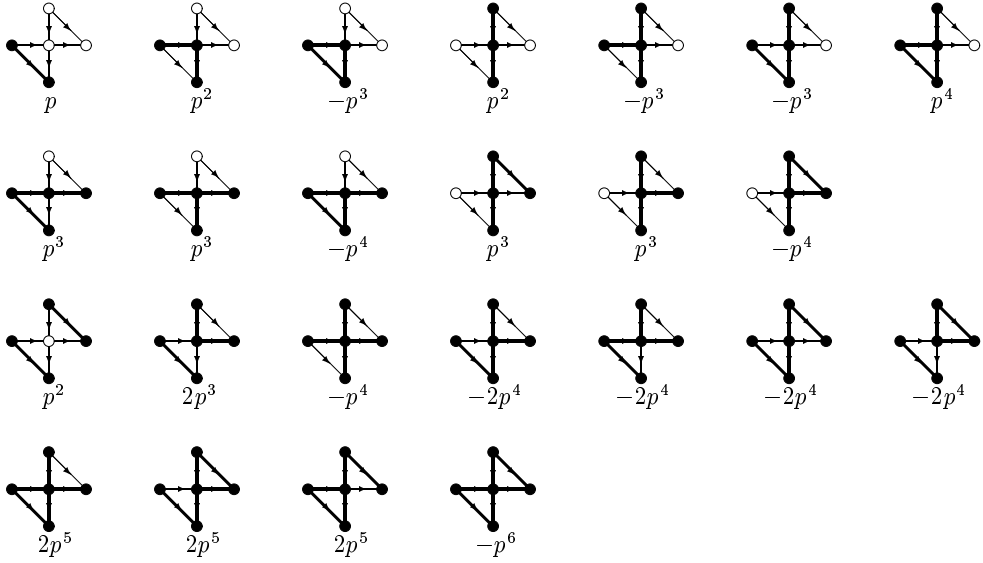}
\end{center}
\caption{\label{fig:kagupdate}
The allowed configurations of occupied edges (thick lines) and their
associated weights generated as the boundary line is moved. The top line 
shows the configurations resulting in the target state $\overline{S}_{1,0}$, 
while the remaining three lines show the configurations leading to 
$\overline{S}_{1,1}$.
} 
\end{figure}

The right panel of figure~\ref{fig:kagtransfer} illustrates how the kink in
the boundary line is moved in order to add an extra cell to the kagom\'e
lattice. The updating rules are derived by considering all possible bond 
configurations and discarding those that do not conform to the basic
rules `vertices with incoming bonds must have outgoing bonds' and `vertices
without incoming bonds have no outgoing bonds'. The weight of each 
allowed configuration is simply calculated by multiplying by a factor
$p$ for each inserted bond and a factor $-1$ for each new cycle. In
 figure~\ref{fig:kagupdate} we show the allowed configurations (and
their associated weights) occurring in the derivation of the updating rules
for the target boundary configurations $\overline{S}_{1,0}$ (top line)
and $\overline{S}_{1,1}$ (remaining lines). The updating rule for the
target $\overline{S}_{0,1}$ follows from symmetry. Summing over all
contributions we get the following set of equations:

\begin{eqnarray}\label{eq:kagbondupdate}
W(\overline{S}_{1,1}) & = & (2p^3-p^4)[W(S_{1,0})+ W(S_{0,1})]+ 
                            (p^2+2p^3-9p^4+6p^5-p^6)W(S_{1,1}), \nonumber \\
W(\overline{S}_{0,1}) & = & p^2W(S_{1,0}) + (p+p^2-p^3)W(S_{0,1}) 
                            - (2p^3-p^4)W(S_{1,1}),  \\
W(\overline{S}_{1,0}) & = & (p+p^2-p^3)W(S_{1,0})+p^2W(S_{0,1})
                          - (2p^3-p^4)W(S_{1,1}), \nonumber \\
W(\overline{S}_{0,0}) & = & W(S_{0,0}). \nonumber
\end{eqnarray}
\noindent
While these equations hold in the general case there is one special case
to consider. When all vertices except those in the kink are in state `0'
the updating is different because we have to carefully avoid forming nodal
points. Thus the only input and output configuration with non-zero weight
is the one in which both vertices in the kink are in state `1'. In this
special case the third configuration in the third row of 
figure~\ref{fig:kagupdate} is forbidden because the central vertex would
be a nodal point and we thus get the updating rule

$$
W(\overline{S}_{1,1}) = (p^2+2p^3-8p^4+6p^5-p^6)W(S_{1,1}).
$$

As for the triangular case we have to use a small algorithm to calculate
$N_{\rm add}$. While the details are slightly different, the general 
considerations are the same in the two cases and we will therefore
refrain from further elaboration. We were able to calculate $C^*_{a,a}$
on rectangles up to width 42 and obtain the series to order 173.

\subsubsection{The concatenation operations \label{sec:kagconcat}}

As indicated above the pair-connectedness is obtained by adding the 
contributions $C_i$ from graphs terminating at a vertex of type $i$.
$C_i$ is in turn obtained by repeated concatenations of non-nodal
graphs. The sequence of concatenations can formally be expressed as:

\begin{equation} \label{eq:kagconcat}
C_i^{(k)} = \sum_j C_j^{(k-1)}C^*_{j,i},
\end{equation}
\noindent
where we start with the initial conditions $C_a^{(0)}=1$,$C_b^{(0)}=0$, and
$C_c^{(0)}=0$. $C_i^{(k)}$ is the contribution after exactly $k$ iterations 
of eq.~(\ref{eq:kagconcat}), $C^*_{j,i}$ is the non-nodal pair-connectedness
from a vertex of type $j$ to one of type $i$, and $C_i$ is finally obtained 
by summation over $k$. As shown above $C^*_{a,a}$ is calculated directly by
the transfer matrix algorithm while the general cases of $C^*_{j,i}$ are
derived from  $C^*_{a,a}$ as follows

\begin{eqnarray}\label{eq:kaggennodal}
C^*_{a,b} & = & tx[p +(p-p^2)C^*_{a,a}], \nonumber \\
C^*_{a,c} & = & tx^{-1}[p +(p-p^2)C^*_{a,a}], \nonumber \\
C^*_{b,a} & = & C^*_{a,b}, \nonumber \\
C^*_{b,b} & = & t^2x^2(p^2-2p^3+p^4)C^*_{a,a}, \\
C^*_{b,c} & = & t^2[p-p^3+(p^2-2p^3+p^4)C^*_{a,a}], \nonumber \\
C^*_{c,a} & = & C^*_{a,c}, \nonumber \\
C^*_{c,b} & = & C^*_{b,c}, \nonumber \\
C^*_{c,c} & = & t^2x^{-2}(p^2-2p^3+p^4)C^*_{a,a}. \nonumber
\end{eqnarray}

In these expressions there are generally two terms representing either `elementary'
steps or more complex non-nodal graphs derived by simple `decorations' of
graphs contributing to $C^*_{a,a}$. As an example we look at $C^*_{a,b}$,
that is non-nodal graphs from black to shaded vertices in fig.~\ref{fig:kagtransfer}.
The first term, $txp$, is just a single horizontal bond, while the second
term, $tx(p-p^2)C^*_{a,a}$, arises as a decoration of a non-nodal graph
from black to black vertices by appending a little triangle to the
bottom right corner of the original non-nodal graph so as to extend
the graph to the nearest shaded vertex to the right. We have to insert
the horizontal bond from the black to the shaded vertex and the diagonal
bond from the white to the shaded vertex. However, we can now either
delete the existing vertical bond from the white to the black vertex,
(yielding an overall additional factor $p$) or leave the bond in place
(yielding the additional factor $-p^2$ since an additional cycle was formed). 
The other expressions involving vertices of type $a$ are derived in the
same manner and correspond to different ways of placing the elementary
steps and the decorating triangle. $C^*_{b,b}$ and $C^*_{c,c}$ involves
decorating a non-nodal graph contributing to $C^*_{a,a}$ by two little
triangles, either at the top pointing left and at the bottom pointing
right or at the top pointing up and at the bottom pointing down. Finally,
$C^*_{b,c}$ contains the two elementary contributions of a diagonal
bond, $t^2p$, or a little triangle, $-t^2p^3$. Note that the simple linear 
graph consisting of a horizontal and vertical bond isn't non-nodal and its
contribution is in fact derived from the concatenation $C^*_{b,a}C^*_{a,c}$.  
The remainder of $C^*_{b,c}$ is obtained by decorating graphs in $C^*_{a,a}$ by 
appending two little triangles one on the top pointing left and one at the bottom 
pointing down.

We could chose not to distinguish between sites of type $b$ and $c$, which
would lead to fewer equations but also make the concatenations less transparent.

\subsubsection{The site problem \label{sec:kagsite}}

Again the updating rules for the site case are derived from the bond rules by 
changing the weights in eq.~(\ref{eq:kagbondupdate}) so as to count the number
of additional occupied vertices rather than edges. This leads to the site updating rules:

\begin{eqnarray}\label{eq:kagsiteupdate}
W(\overline{S}_{1,1}) & = & p^3[W(S_{1,0})+ W(S_{0,1})]+ 
                            (p^2-2p^3)W(S_{1,1}), \nonumber \\
W(\overline{S}_{0,1}) & = & p^2W(S_{1,0}) + pW(S_{0,1}) 
                            - p^2W(S_{1,1}),  \\
W(\overline{S}_{1,0}) & = & pW(S_{1,0})+p^2W(S_{0,1})
                          - p^2W(S_{1,1}), \nonumber \\
W(\overline{S}_{0,0}) & = & W(S_{0,0}). \nonumber
\end{eqnarray}
\noindent

The concatenation is performed as in the bond case. However, as we changed
the updating rules so we have to change $C^*_{i,j}$. A little calculation will
show that  the terms $C^*_{i,j}$ in the site case are very simple and are derived 
from eq.~(\ref{eq:kaggennodal}) by deleting all factors involving $C^*_{a,a}$,
leaving just the elementary steps from $a$ to $b$ and so on. 

We calculated $C^*_{a,a}$ on rectangles up to width 41 and thus 
obtained the series to order 169.

\subsection{(4.8$^2$) lattice \label{sec:4_8calc}}

In the case of the $(4.8^2)$-lattice we define a slightly restricted
pair-connectedness, which in essence corresponds to a special choice
for the initial vertex and a restriction on the possible terminal bonds.
The details of our choice of initial vertex should be unimportant as
long as we preserve the symmetry among the horizontal and vertical
directions. We have chosen to introduce a special initial vertex as
indicated in the left panel of figure~\ref{fig:4_8transfer}. This vertex 
is, in the parlance of electrical networks, the point at which current
is injected into the network and the current may flow along either
one or both of the adjoining horizontal and vertical edges. Bonds
along the short edges connecting the initial vertex to the network
are not counted in $|g|$. Finally, for our convenience, we have
chosen to specify that we always terminate the flow in either a horizontal 
or vertical bond. In other words $C_{t,x}$ is the probability of finding a 
path from the initial vertex to either a horizontal or vertical edge.

\subsubsection{The transfer-matrix algorithm \label{sec:4_8transfer}}

\begin{figure}
\begin{center}
\includegraphics[scale=0.8]{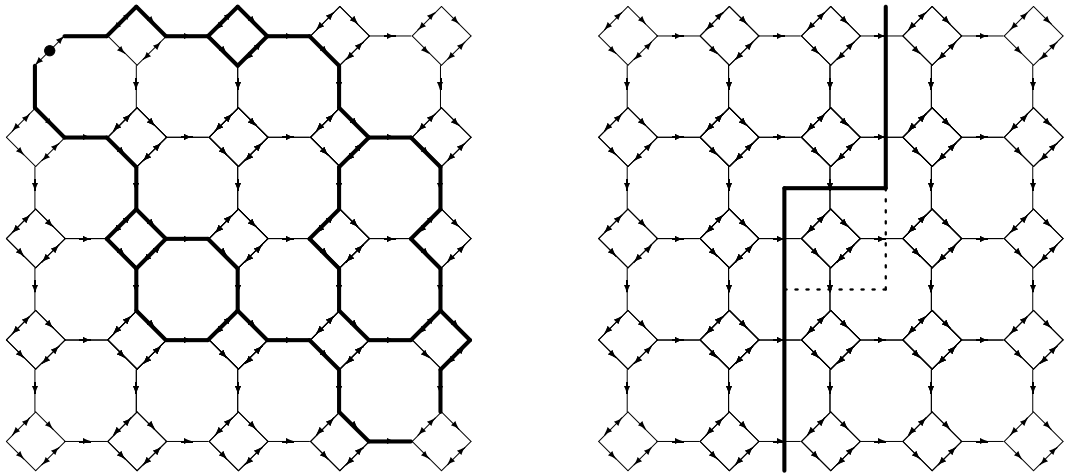}
\end{center}
\caption{\label{fig:4_8transfer}
A piece of the directed $(4.8^2)$ lattice. Shown in the left panel is 
an example of a non-nodal graph. The right panel illustrates how the 
boundary line is moved during the transfer-matrix calculation.
} 
\end{figure}

The transfer matrix algorithm is designed to calculate the contribution
from non-nodal graphs starting and terminating on adjoining horizontal and 
vertical edges, as illustrated in the left panel of 
figure~\ref{fig:4_8transfer}. Again we calculate $C^*_{t,x}$ for graphs
on rectangles of size $w \times l$, and as usual we need only consider
rectangles with $l \geq w$. These graphs contribute to $C^*_{t,x}$ for
$t=l+w$ and $x=l-w$, and contain at least $2w+6l=4t+2x$ bonds. Thus,
in a calculation or order $N$, we must calculate the contributions from
rectangles of widths up to $N/8$ and lengths from $w$ to $(N-8w)/6$.

\begin{figure}
\begin{center}
\includegraphics[scale=0.8]{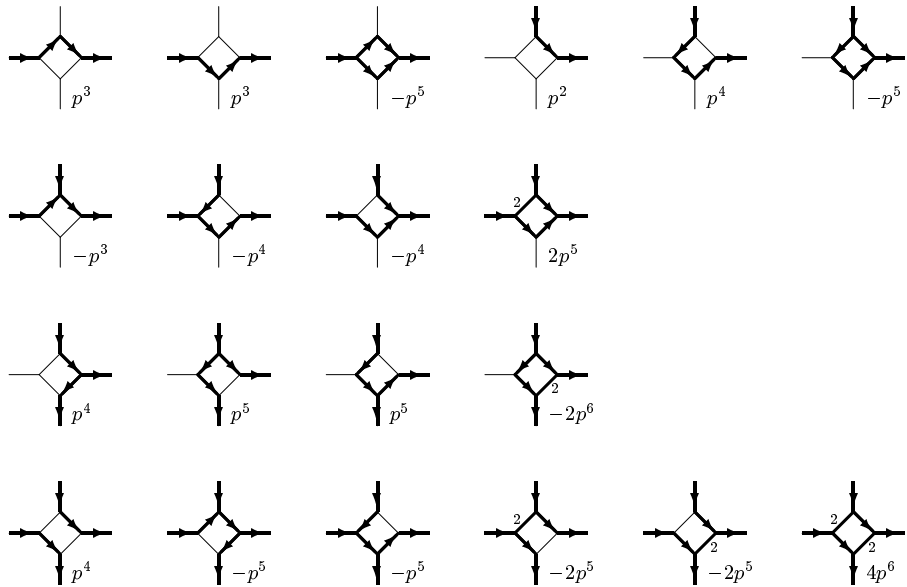}
\end{center}
\caption{\label{fig:4_8update}
The allowed configurations of occupied edges (thick lines) and their
associated weights generated as the boundary line is moved. The top two lines 
show the configurations resulting in the target state $\overline{S}_{0,1}$, 
while the remaining two lines show the configurations leading to 
$\overline{S}_{1,1}$. In each case we show the direction of the flow.
Along the thick edges with no arrows the flow can progress in either
direction, and hence we have marked these edges with a `2' indicating that
there are two distinct flow patterns.
} 
\end{figure}

The right panel of figure~\ref{fig:4_8transfer} shows how the boundary line
is moved in order to add an extra cell to the $(4.8^2)$-lattice. The 
updating rules are derived in a similar fashion to those for the kagom\'e 
lattice. The only major difference is that along some of the diagonal
edges the current can flow in either direction, so in some cases depending
on the configuration of incoming and outgoing bonds we may have several
distinct flow patterns associated with a given bond configuration. In
figure~\ref{fig:4_8update} we show the configurations and the associated
weights occurring in the derivation of the updating rules for the target
configurations $\overline{S}_{0,1}$ (top two lines) and
$\overline{S}_{1,1}$ (remaining two lines). This leads to the following
set of equations

\begin{eqnarray}\label{eq:4_8bondupdate}
W(\overline{S}_{1,1}) & = & (p^4+2p^5-2p^6)[W(S_{1,0})+ W(S_{0,1})]+ 
                            (p^4-6p^5+4p^6)W(S_{1,1}), \nonumber \\
W(\overline{S}_{0,1}) & = & (2p^3-p^5)W(S_{1,0}) + (p^2+p^4-p^5)W(S_{0,1}) 
                            - (p^3+2p^4-2p^5)W(S_{1,1}),  \\
W(\overline{S}_{1,0}) & = & (p^2+p^4-p^5)W(S_{1,0})+(2p^3-p^5)W(S_{0,1})
                          - (p^3+2p^4-2p^5)W(S_{1,1}), \nonumber \\
W(\overline{S}_{0,0}) & = & W(S_{0,0}). \nonumber
\end{eqnarray}
\noindent

As for the previous cases we wrote an algorithm to calculate $N_{\rm add}$
and again the same general considerations apply. In this case we were able
to calculate $C^*_{t,x}$ on rectangles up to width 31 and consequently
calculate the series to order 255.

\subsubsection{The concatenation operations \label{sec:4_8concat}}

The restricted pair-connectedness is calculated by adding the contributions
$C_h$ and $C_v$ from graphs terminating with a horizontal or vertical
bond, respectively. As in the previous cases these contributions are
obtained from repeated concatenations

\begin{eqnarray}\label{eq:4_8bondconcat}
C_h^{(k)} & = & C_h^{(k-1)} C_{h,h}^* + C_v^{(k-1)} C_{v,h}^*, \nonumber \\
C_v^{(k)} & = & C_h^{(k-1)} C_{h,v}^* + C_v^{(k-1)} C_{v,v}^*,  
\end{eqnarray}
\noindent
where we start with the initial condition
\begin{eqnarray}
C_h^{(0)} & = & txp+tx(-p^3-2p^4+2p^5) C^*, \nonumber \\
C_v^{(0)} & = & tx^{-1}p+tx^{-1}(-p^3-2p^4+2p^5) C^*,  
\end{eqnarray}
\noindent
and $C^*_{i,j}$ is the non-nodal pair-connectedness from an edge of type
$i$ to one of type $j$. As described above $C^*$ is calculated using the
transfer matrix algorithm, while $C^*_{i,j}$ are easy to derive from the
updating rules in eq.~(\ref{eq:4_8bondupdate})

\begin{eqnarray}
C_{h,h}^* & = & tx[2p^3-p^5-p^5(1+2p-2p^2)^2 C^*], \nonumber \\
C_{h,v}^* & = & tx^{-1}[p^2+p^4-p^5-p^5(1+2p-2p^2)^2 C^*], \nonumber \\
C_{v,h}^* & = & tx[p^2+p^4-p^5-p^5(1+2p-2p^2)^2 C^*],  \\
C_{v,v}^* & = & tx^{-1}[2p^3-p^5-p^5(1+2p-2p^2)^2 C^*]. \nonumber 
\end{eqnarray}

The pre-factor multiplying $C^*$ arise as the product of two contributions
namely going from a single occupied bond to two occupied bonds and the
reverse. The weights of each of these contributions can be extracted from  
eq.~(\ref{eq:4_8bondupdate}). The first contribution is derived from 
the updating rule for $W(\overline{S}_{1,1})$ as generated from either
$W(S_{1,0})$ or $W(S_{0,1})$. Noting that two of the bonds 
inserted in this updating operation are already counted as part of the 
non-nodal graph we find the weight $p^2(1+2p-2p^2)$. The second contribution
is derived from the updating rule for $W(\overline{S}_{1,0})$ 
or $W(\overline{S}_{0,1})$ as generated from $W(S_{1,1})$ and thus carries 
the weight $-p^3(1+2p-2p^2)$, thus giving us the total pre-factor
$-p^5(1+2p-2p^2)^2$. In addition we have to add the weights corresponding
to simple graphs not involving non-nodal pieces. These contributions arise
from the graphs shown on the first line in fig.~\ref{fig:4_8update}.
Note that strictly speaking these graphs are not non-nodal but they
are the simplest graphs not counted in $C^*$ and they are non-nodal because
we have chosen only to allow graphs to terminate at horizontal or vertical
edges.

\subsubsection{The site problem \label{sec:4_8site}}

Changing the weights in eq.~(\ref{eq:4_8bondupdate}) by counting the number
of additional occupied vertices rather than edges leads to the site 
following updating rules:

\begin{eqnarray}\label{eq:4_8siteupdate}
W(\overline{S}_{1,1}) & = & p^3[W(S_{1,0})+ W(S_{0,1})]- 
                            p^4W(S_{1,1}), \nonumber \\
W(\overline{S}_{0,1}) & = & (2p^3-p^4)W(S_{1,0}) + p^2W(S_{0,1}) 
                            - p^3W(S_{1,1}),  \\
W(\overline{S}_{1,0}) & = & p^2W(S_{1,0})+(2p^3-p^4)W(S_{0,1})
                          - p^3W(S_{1,1}), \nonumber \\
W(\overline{S}_{0,0}) & = & W(S_{0,0}). \nonumber
\end{eqnarray}
\noindent

The concatenation operations (\ref{eq:4_8bondconcat}) are identical but the initial conditions
are 
\begin{eqnarray}
C_h^{(0)} & = & txp-txp^2 C^*, \nonumber \\
C_v^{(0)} & = & tx^{-1}p-tx^{-1}p^2 C^*,  
\end{eqnarray}
\noindent
while
\begin{eqnarray}
C_{h,h}^* & = & tx[2p^3-p^4-p^4 C^*], \nonumber \\
C_{h,v}^* & = & tx^{-1}[p^2-p^4 C^*], \nonumber \\
C_{v,h}^* & = & tx[p^2-p^4 C^*],  \\
C_{v,v}^* & = & tx^{-1}[2p^3-p^4-p^4 C^*]. \nonumber 
\end{eqnarray}

In this case we calculated $C^*_{t,x}$ on rectangles up to width 29 and consequently
the series to order 239.

\section{Analysis of the series \label{sec:ana}}

The various series were analysed using inhomogeneous differential 
approximants \cite{AJG89a}. Our use of differential approximants 
for series analysis has been detailed in previous papers 
\cite{IJ99a,IJ96a} and the interested reader can refer
to these papers and the comprehensive review \cite{AJG89a}
for further details. Suffice to say that this technique of series analysis 
usually enables one to obtain  accurate estimates for the critical 
point $p_c$, the associated critical exponents, and possible 
non-physical singularities. 

In the following sections we shall briefly describe the results obtained
from the series analysis.

\subsection{Critical points and exponents \label{sec:anacrit}}

In table~\ref{tab:critical} we have listed the estimates for
the critical points and exponents for the problems studied in this
paper as obtained by analyzing the series for the average cluster
size and the first non-zero parallel and perpendicular moments of the 
pair-connectedness. The estimates for $p_c$ and the critical exponents
are based on results of series analysis using both second and third 
order approximants with varying degrees of the inhomogeneous polynomial. 
The estimates were obtained by averaging values obtained from 
second and third order differential approximants. For each order $L$ of 
the inhomogeneous polynomial we averaged over those approximants to the 
series which used  at least the first $90\%$ of the terms in the series.
The error quoted for these estimates reflects the spread among the 
approximants. Note that these error bounds should
{\em not} be viewed as a measure of the true error as they cannot include
possible systematic sources of error.
For completeness and comparison we have also listed the results for the 
square lattice bond and site problems from \cite{IJ99a}. Note that in this case 
the exponent estimates are not necessary obtained by direct analysis
of the corresponding series.

\begin{table}
\caption{\label{tab:critical}
Estimates for $p_c$ and critical exponents for various
directed percolation problems.}
\begin{tabular}{lllll} \hline 
Problem & $p_c$ & $\gamma$ & $\gamma+\nu_{||}$ & $\gamma+2\nu_{\perp}$\\ \hline
Square bond & 0.644700185(5) & 2.277730(5) & 4.011577(11) & 4.471438(13)  \\
Square site & 0.70548515(20) & 2.27765(6) & 4.01135(15) & 4.47130(8)  \\
Honeycomb bond & 0.82285680(6) & 2.27765(5) & 4.0113(1) & 4.47135(5) \\
Triangular bond & 0.47802525(5) & 2.27783(6) & 4.01176(8) & 4.47156(6) \\
Triangular site & 0.59564675(10) & 2.27771(2) & 4.01145(5) & 4.47144(2) \\
Kagom\'e bond & 0.65896910(8) & 2.27790(6) & 4.01186(8) & 4.47167(9) \\
Kagom\'e site & 0.73693182(4) & 2.27778(3) & 4.01160(5) & 4.47150(4) \\
$(4.8^2)$ bond & 0.76765116(7) & 2.27765(4) & 4.01144(3) & 4.47128(6) \\
$(4.8^2)$ site & 0.81674220(4) & 2.27770(6) & 4.01145(4) & 4.47127(4) \\  
\hline  \hline
\end{tabular} 
\end{table}

In all cases we have obtained accurate estimates of the critical point
$p_c$ and the critical exponents, with the exponent estimates differing 
only in the $5^{th}$ digit. In most cases we notice that the error bounds are 
an order of magnitude larger than for the square bond case. Standard 
arguments from renormalization group theory says that the exponents should 
be universal. This expectation is clearly supported by the present analysis. 
Obviously, the exponent estimates for some of the problems do not quite
agree with the square bond case within the quoted error bounds. However,
the differences are quite small and any discrepancy is likely to be due
to the difficulty of obtaining reliable error bounds. Certainly the
discepancies are not of such an order as to challenge the strong
belief in universality. In particular we note that the exponent estimates
scatter around the square bond value, with some being higher (triangular
bond, kagom\'e bond and site) others being lower (square site, honeycomb bond,
$(4.8^2)$ bond) and yet others being in complete agreement (triangular
site, $(4.8^2)$ site). Futhermore, we note that in many cases (e.g. honeycomb
bond) if we look at other exponent estimates, say the one we can
obtain of $2\nu_{\perp}$, there is a close agreement with the square bond
case. The square and honeycomb bond cases yield $2\nu_{\perp}=2.193708(18)$
and $2\nu_{\perp}=2.19370(11)$, respectively. The worst case scenario therefore
would be that we would have to adopt a more conservative error estimate
on the exponents. For reasons explained in detail in \cite{IJ99a} we have
a great deal of confidence in the square bond estimates and are reluctant
to increase the error bounds. More cautious readers may choose to take a 
different view.

\begin{figure}
\begin{center}
\includegraphics[scale=0.8]{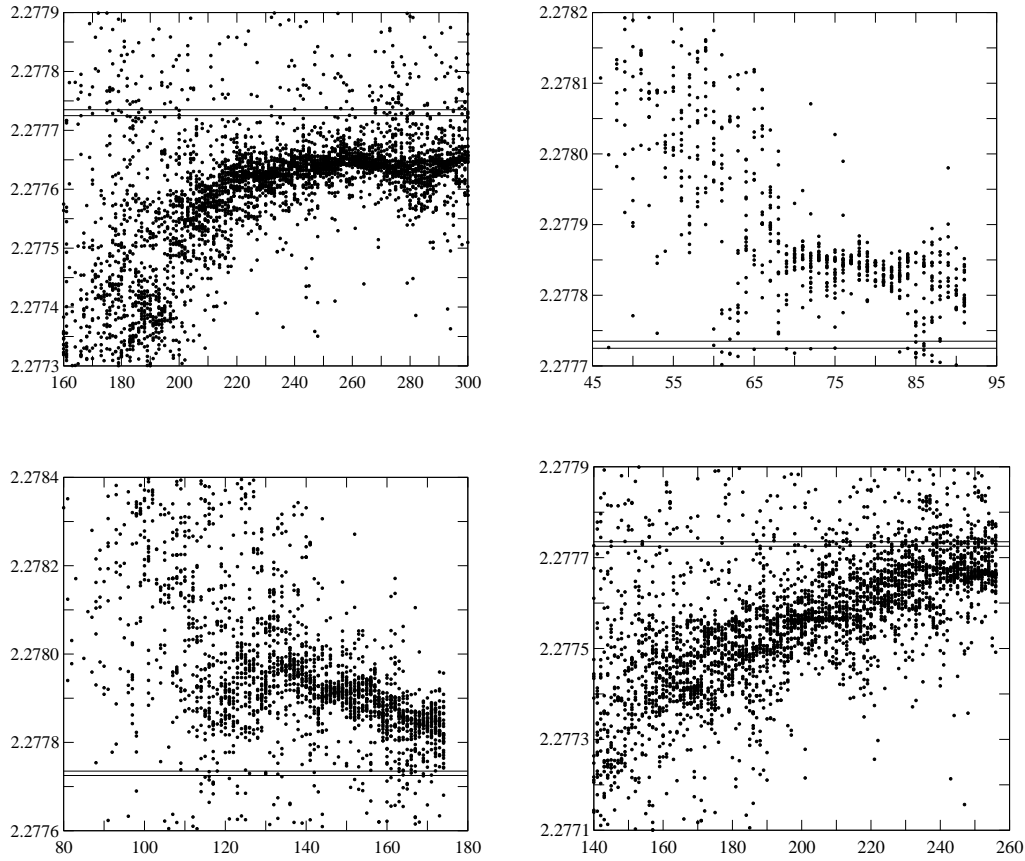}
\end{center}
\caption{\label{fig:csexpnt}
Estimates of the critical exponent $\gamma$ vs. the number of terms used 
by third order differential approximants to the cluster-size series $S(p)$ 
for bond percolation on the directed honeycomb (top left), triangular (top right), 
kagom\'e (bottom left), and $(4.8^2)$ (bottom right) lattices.
} 
\end{figure}

As we have already mentioned above, one of the major unresolved problems of 
series analysis is the calculation of reliable error estimates. So in trying to 
confirm, as we are here, the universality of the critical exponents, it is often 
useful to plot the behaviour of the exponent estimates versus the number of terms 
used by the differential approximants. In this way it often possible to gauge more 
clearly whether or not the high-order approximants have settled down to the limiting 
value of the true exponent. In fig.~\ref{fig:csexpnt} we carry out such an analysis 
of the cluster size series for the directed bond percolation problems. 
Each point in the four panels correspond to an estimate of $\gamma$, obtained 
from a third order differential approximant, plotted against the 
number of terms used by the differential approximant. The straight
lines indicate the error bounds on the very accurate estimate of 
$\gamma$ obtained from the analysis of the square bond series.
From these plots we can see that the estimates of $\gamma$ obtained
from the triangular (top right panel) and kagome\'e lattices (bottom left panel)
exhibit a pronounced downwards drift as the number of terms is increased,
while the estimates from the $(4.8^2)$ lattice (bottom right panel) 
display an upwards drift. So the estimates of $\gamma$ have
not yet settled at their limiting value, but there can be no
doubt that the indicated value of $\gamma$ is fully consistent
with the estimates. The only possible disagreement comes from the honeycomb
problem (top left panel).

\subsection{Non-physical singularities \label{sec:ananps}}

Non-physical singularities are of interest for several reasons.
Firstly, one expects that the presence of non-physical
singularities (particularly if they are closer to the origin than
$p_c$) can have a dramatic influence on the precision of the estimates 
of $p_c$ and the critical exponents. Secondly,  knowledge of the position 
and associated exponents of non-physical singularities may help in the search
for exact solutions and in some cases one may gain a better understanding
of the problem by studying the behaviour of various physical 
quantities as analytic functions of complex variables. 
Many of the series have a radius of convergence smaller than $p_c$ due to 
singularities in the complex $p$-plane closer to the origin than the 
physical critical point. Since all the coefficients in the expansion 
are real, complex singularities always come in conjugate pairs. 
That such non-physical singularities must be present is evident from the fact that
the coefficients in the various series change sign. If only the physical
singularity was present all coefficients would have the same sign.
In the following we shall briefly outline our findings. A more
detailed decription of the method can be found in \cite{IJ99a}.

\paragraph{Honeycomb lattice problem.} The series have a pronounced and well 
defined singularity at $0.022734(2)\pm 0.708909(2)i$. The exponents estimates 
are consistent with the exact values  1/2, $-1/2$, $-3/2$, and $-1/2$ for the 
series $S(p)$, $\mu_{1,0}$, $\mu_{2,0}$, and $\mu_{0,2}$, respectively. 
A second singularity occurs at $0.34715(2)\pm 0.58813(3)i$ with exponent
estimates consistent with the values 3, 2, 1, and 2 for the 
series $S(p)$, $\mu_{1,0}$, $\mu_{2,0}$, and $\mu_{0,2}$, respectively.
In addition there are other less well defined singularities
for which no meaningful exponent estimates can be obtained.
We estimate the positions of these singularities to be
$-0.6183(2)$ and $-0.180(5)\pm 0.725(5)i$.

\paragraph{Triangular lattice problems.} The series do not
appear to have any well defined non-physical singularities. There
is some evidence of  singularities for the site problem
at $-0.323(5)\pm 0.39(1)i$.

\paragraph{Kagom\'e lattice problems.} The series for the bond problem shows 
no evidence of clearly defined non-physical singularities. 
The series for
the site problem have a singularity on the negative axis at $-0.61804(2)$
with exponents $-0.344(3)$, $-1.34(1)$, $-2.34(1)$ and $-1.656(5)$ for the 
series $S(p)$, $\mu_{1,0}$, $\mu_{2,0}$, and $\mu_{0,2}$, respectively.
The is evidence for further singularities at $-0.53(1)\pm 0.375(5)i$

\paragraph{$(4.8^2)$ lattice problems.} The series for bond problem have
a singularity at $-0.61804(2)$
with exponents $-0.505(5)$, $-1.505(5)$, $-2.49(1)$ and $-1.555(10)$ for the 
series $S(p)$, $\mu_{1,0}$, $\mu_{2,0}$, and $\mu_{0,2}$, respectively.
Further singularities occur at $0.428(4)\pm 0.458(4)i$, $0.215(5)\pm 0.590(5)i$,
and $-0.528(5) \pm 0.346(5)i$. The series for the site  problem have
a singularity at $-0.618034(3)$ and also at $0.18536(2)\pm 0.65939(3)i$.
In these cases the exponents are consistent with the
exact values $1/2$, $-1/2$, $-3/2$ and $-1/2$ for the 
series $S(p)$, $\mu_{1,0}$, $\mu_{2,0}$, and $\mu_{0,2}$, respectively.
Singularities also occur at $-0.3000(5)\pm0.6247(2)$ and $0.428(2)\pm 0.536(3)$.

\section{Summary and discussion \label{sec:sum}}

We have presented new algorithms for the calculation of low-density
series expansions for directed percolation on several two-dimensional 
lattices. Analysis of the series yielded accurate estimates of the 
critical point $p_c$ and various critical exponents, including
the exponent $\gamma$ governing the divergence of the cluster size
at $p_c$. The estimates of the critical exponents confirm, to a
high degree of accuracy and confidence, the universality of the
exponents for directed percolation.  This conclusion is in full
agreement with  earlier studies on the triangular \cite{EGDB88a,IJ96a}, 
honeycomb \cite{JG96} and kagom\'e lattices \cite{RC91} which also
confirm universality although with lower precision.

To some extent the exponent estimates in Table~\ref{tab:critical} do not
agree with the very accurate square bond estimates. This
is largely explained by the plots in Fig.~\ref{fig:csexpnt}, which shows
that in many cases the exponent estimates exhibit a pronounced
drift and thus have not yet settled down to their limiting value.
A further complicating factor is the presense of several non-physical
singularities. They appear to be particularly prominent in the
series for the honeycomb bond problem, which might explain
the slight discrepancy between the exponent estimates for this
problem and those for the square bond problem.

\section*{E-mail or WWW retrieval of series}

The series for the directed percolation problems studied in this paper 
can be obtained via e-mail by sending a request to 
I.Jensen@ms.unimelb.edu.au or via the world wide web on the URL
http://www.ms.unimelb.edu.au/\~{ }iwan/ by following the relevant links.

\section*{Acknowledgments}

The calculations presented in this paper were in part performed on
the facilities of the Australian Partnership for Advanced Computing (APAC) 
and the Victorian Partnership for Advanced Computing (VPAC). 
We gratefully acknowledge financial support from the Australian Research Council.


\begin{thebibliography}{10}

\bibitem{AE77}
Arrowsmith D~K and Essam J~W 1977 Percolation theory on directed graphs {\em J.
  Math. Phys.\/} {\bf 18} 235--238

\bibitem{BE84}
Bhatti F~M and Essam J~W 1984 The resistive susceptibility of random
  diode-insulator networks {\em J. Phys. A: Math. Gen.\/} {\bf 17} L67--L73

\bibitem{DPB82}
Dhar D, Phani M~K and Barma M 1982 Enumeration of directed site animals on
  two-dimensional lattices {\em J. Phys. A: Math. Gen.\/} {\bf 15} L279--L284

\bibitem{Essam80a}
Essam J~W 1980 Percolation theory {\em Rep. Prog. Phys.\/} {\bf 43} 833--912

\bibitem{EDB82}
Essam J~W and De'{B}ell K 1982 The pair connectedness for directed percolation
  on the honeycomb and diamond lattices {\em J. Phys. A: Math. Gen.\/} {\bf 15}
  L601--L604

\bibitem{EGDB88a}
Essam J~W, Guttmann A~G and De'Bell K 1988 On two-dimensional directed
  percolation {\em J. Phys. A: Math. Gen.\/} {\bf 21} 3815--3832

\bibitem{AJG89a}
Guttmann A~J 1989 Asymptotic analysis of power-series expansions in {\em Phase
  Transitions and Critical Phenomena\/} (eds. C~Domb and J~L Lebowitz) (New
  York: Academic) vol.~13 pp. 1--234

\bibitem{JG96}
Jensen I and Guttmann A~J 1996 Series expansions of the percolation probability
on the directed triangular lattice {\em J. Phys. A: Math. Gen.\/} {\bf 29} 497--517

\bibitem{IJ96a}
Jensen I 1996 Low-density series expansions for directed percolation on square
  and triangular lattices {\em J. Phys. A: Math. Gen.\/} {\bf 29} 7013--7040

\bibitem{IJ99a}
Jensen I 1999 Low-density series expansions for directed percolation {I.} {A}
  new efficient algorithm with applications to the square lattice {\em J. Phys.
  A: Math. Gen.\/} {\bf 32} 5233--5249

\bibitem{RC91}
Rushin H~J and Cadilhe A 1991 Percolation on directed Archimedean nets 
{\em J. Non-Crystalline Solids\/} {\bf 127} 114--120


\bibitem{StaufferPercBook}
Stauffer D and Aharony A 1992 {\em Introduction to Percolation Theory\/}
  (London: Taylor \& Francis) 2 ed.

\end{thebibliography}
\end{document}